\documentstyle[twoside,fleqn,espcrc2,epsf]{article}

\def\lsim{\raise0.3ex\hbox{$<$\kern-0.75em\raise-1.1ex\hbox{$\sim$}}}
\def\gsim{\raise0.3ex\hbox{$>$\kern-0.75em\raise-1.1ex\hbox{$\sim$}}}

\newcommand{\pslash}{p\kern-1ex /}
\newcommand{\Dslash}{{\cal D}\kern-1.5ex /}

\newcommand{\beqa}{\begin{eqnarray}}
\newcommand{\eeqa}{\end{eqnarray}}

\newcommand{\beq}{\begin{equation}}
\newcommand{\eeq}{\end{equation}}
\newcommand{\bc}{\begin{center}}
\newcommand{\ec}{\end{center}}

\title{\vspace*{-2.cm}
\begin{flushright}
{\normalsize UTHEP-460}\\
{\normalsize UTCCP-P-125}\\
\end{flushright}
Study of unstable particle through the spectral function
in $O(4)$ $\phi^4$ theory\thanks{Talk presented by T.\ Yamazaki}}
\author{
  N.~Ishizuka\address{Institute of Physics,
    University of Tsukuba, Tsukuba, Ibaraki 305-8571, Japan}$^{\rm ,}$\address{Center for Computational Physics, University of Tsukuba, Tsukuba, Ibaraki 305-8577, Japan}
  and 
  T.~Yamazaki$^{\rm a}$
}

\begin{document}
\pagestyle{empty}

\begin{abstract}
We test application of the maximum entropy method
to decompose the states 
contributing to the unstable $\sigma$ correlation function
through the spectral function
in the four dimensional $O(4)$ $\phi^4$ theory.
Reliable results are obtained for the $\sigma$ mass and 
two-particle $\pi\pi$ state energy
using only the $\sigma$ correlation function.
We also find that the property of the $\sigma$ particle is
different between the unstable ($m_{\sigma}/m_{\pi}>2$) and
stable ($m_{\sigma}/m_{\pi}<2$) cases.
\end{abstract}

\maketitle

\section{Introduction}
\label{sec:intro}
The decay of particles, {\it e.g., } $\rho$ meson decay 
is not understood well on the lattice. 
One of the difficulties is that 
an unstable particle and its intermediate states, such 
as the $\rho$ meson and the $\pi\pi$ state
in the isospin-1 channel, have the same quantum numbers, 
leading to a multi-exponential form of the unstable particle
correlation function. 
To obtain the energies of the states
we have to decompose the states in the correlation function.

In a finite volume 
it is possible to decompose the states
with the spectral function, which can be 
numerically reconstructed
from the correlation function
with the maximum entropy method
(MEM).
MEM has been recently applied to meson correlation functions
in quenched lattice QCD~\cite{my}.
Here we explore application of MEM to calculate the energies of states in 
unstable particle systems.


\section{Model and parameters}
Since the calculation of the unstable particle is difficult in QCD, 
we use the four dimensional $O(4)$ $\phi^4$ theory
in this work.
In this theory the $\sigma$ particle and two-pion state  
in the isospin-0 channel have the same quantum numbers.
These states overlap in the $\sigma$ correlation function.

We calculate the correlation functions for $\sigma$ and 
$\pi\pi$ with zero momentum for the two cases, one 
where $\sigma$ is unstable ($m_{\sigma}/m_{\pi}\approx 3.7$) and
the other stable ($m_{\sigma}/m_{\pi}\approx 1.8$). 
Several spatial lattice sizes in the range $10^3-28^3$ are employed 
to study the volume dependences of the spectral functions, while 
the temporal lattice size is fixed to 64. 
We perform 0.6 to 1.2 million iterations per simulation point
and lattice size.
The method of simulation and measurements of the correlation
functions follow ref.~\cite{O4}.
We calculate $\langle (O(\tau)-O(\tau+1))O(0)\rangle$ to
subtract the vacuum contribution where $O$ denotes $\sigma$ or $\pi\pi$ 
operator.

\section{Results}
\subsection{Spectral function data}

\begin{figure}[t!]
\vspace{-.5cm}
\centerline{\epsfxsize=7.4cm \epsfbox{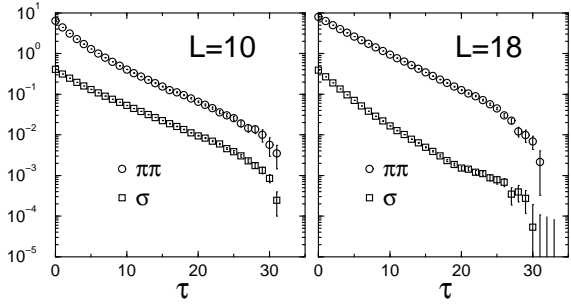}}
\vspace{-1.3cm}
\caption{Correlation functions for the $\sigma$ and
$\pi\pi$ in the unstable case.
\label{fig:corr}}
\vspace{-8mm}
\end{figure}

In Fig.\ref{fig:corr}
we illustrate the $\sigma$ and $\pi\pi$ correlation functions 
at two volumes in the unstable case.
For smaller volume the slopes of the two correlation functions agree, 
while for larger volume the slopes are different 
and the $\sigma$ correlation function
is of multi-exponential form.
This means that the overlap of the $\pi\pi$ state 
dominates the $\sigma$ correlation function for small volume, 
while it decreases as volume increases.

\begin{figure}[t!]
\vspace{-.5cm}
\centerline{\epsfxsize=7cm \epsfbox{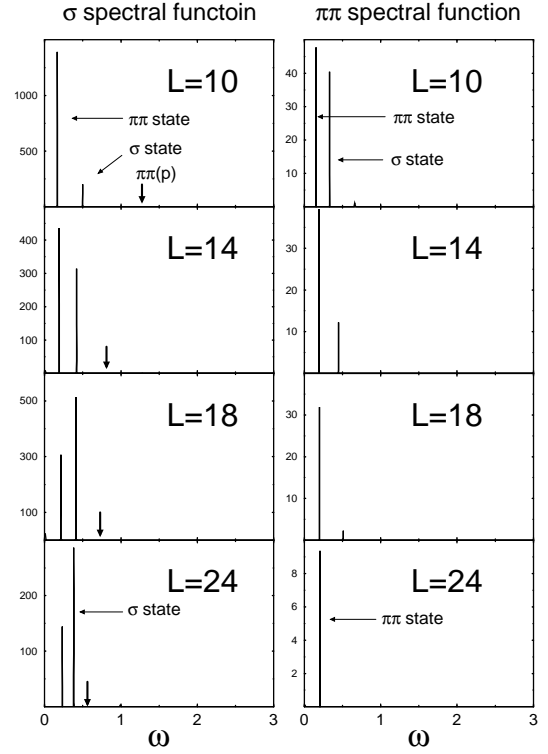}}
\vspace{-1.3cm}
\caption{Spectral functions for the $\sigma$ and
$\pi\pi$ in the unstable case.
\label{fig:spec-lat}}
\vspace{-6mm}
\end{figure}

The spectral function reconstructed from the
correlation function at various volumes
in the unstable case is presented in Fig.~\ref{fig:spec-lat}.
The parameters for MEM are compiled in ref.~\cite{myO4}.
We find the spectral functions to be a sum of 
$\delta$-function peaks as expected.

The first peak corresponds to the $\pi\pi$ state with zero momentum
and the second peak to the $\sigma$ state.
As seen from the peak height, the overlap of the $\pi\pi$ state
in the $\sigma$ spectral function decreases with increasing volume. 
This explains the difference of behavior of the $\sigma$
correlation function in Fig.~\ref{fig:corr}.
The $\sigma$ peak in the $\pi\pi$ spectral
function also decreases as the volume increases.

In the $\sigma$ spectral function
the downward arrow marks the position where 
the $\pi\pi$ state with relative momentum $p=2\pi/L$  should appear.
The state is not observed, showing that the overlap is too small
in these small volumes.

\subsection{Sigma mass and two-pion energy}

In the unstable case, the $\sigma$ mass and $\pi\pi$ energy
are obtained from the positions of the peaks in 
both the $\sigma$ and $\pi\pi$ spectral functions.
We compare results from this analysis with
those obtained with the diagonalization of
the correlation function matrix~\cite{ph1}.
The results are shown in Fig.~\ref{fig:ene_308}; 
MEM($\sigma$) and MEM($\pi\pi$) denote data obtained from 
the $\sigma$ and $\pi\pi$ spectral functions, respectively.
For larger volumes the $\sigma$ mass from MEM($\pi\pi$)
and the $\pi\pi$ energy from MEM($\sigma$) suffer from 
large errors or do not appear, 
because the overlaps of the states are small for large volumes.
We find the $\sigma$ mass from MEM to be reasonably
consistent with that from the diagonalization. 
For the $\pi\pi$ energy all results agree well.

\begin{figure}[t!]
\vspace{-.5cm}
\centerline{\epsfxsize=7cm \epsfbox{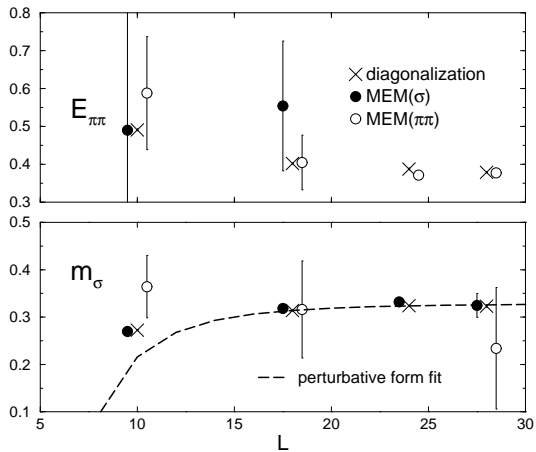}}
\vspace{-1.2cm}
\caption{Energies in the stable case.
\label{fig:ene_30415}}
\vspace{-7mm}
\end{figure}

In the stable case we also obtain the $\sigma$ mass
and $\pi\pi$ energy with MEM and diagonalization.
The results are presented in Fig.~\ref{fig:ene_30415}. 
The two set of data are reasonably consistent.

Comparing Fig.~\ref{fig:ene_308} with Fig.~\ref{fig:ene_30415},
we find that there is an essential difference in the volume dependence
of the $\sigma$ mass and $\pi\pi$ energy
between the unstable and stable cases.
In the unstable case the $\sigma$ mass decreases
as the volume increases
while it increases in the stable case.
The $\pi\pi$ energy also shows a different volume dependence 
for the two cases.

\subsection{Volume dependence of $m_{\sigma}$}

\begin{figure}[t!]
\vspace{-.5cm}
\centerline{\epsfxsize=7cm \epsfbox{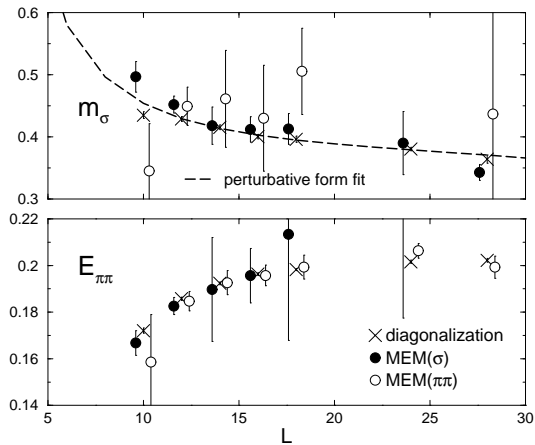}}
\vspace{-1.2cm}
\caption{Energies in the unstable case.
\label{fig:ene_308}}
\vspace{-8mm}
\end{figure}

In order to understand the volume dependence of $\sigma$ mass, 
we fit our data from diagonalization 
assuming the perturbative form in a finite volume given by
$
m_{\sigma}(L) = m_{\sigma} + g_R
(\Delta m_{\sigma}(L)-\Delta m_{\sigma}(\infty))
$
where 
$
\Delta m_{\sigma}(L) =
\frac{1}{4m_{\sigma}L^3}\sum_{\vec{p}}\left[
\frac{D}{W_{\pi}(p)}+\frac{D}{W_{\sigma}(p)}
+2C_{\pi}(p)+6C_{\sigma}(p)\right]
$
with 
$D=
1-\frac{3(m_{\sigma}^2-m_{\pi}^2)}{m_{\sigma}^2}$,
$C_{\alpha}(p) =
\frac{m_{\sigma}^2-m_{\pi}^2}
{W_{\alpha}(p)(m_{\sigma}^2-W_{\alpha}^2(p))},
$
and
$W_{\alpha}(p) = 2\sqrt{m_{\alpha}^2+\hat{p}^2}$, $\hat{p_i}=2\sin(p_i/2)$,
$p_i=2\pi n_i/L$.
Here $m_{\sigma}$ is $\sigma$ mass in the infinite volume and
$g_R$ is the renormalized coupling, which are the fit parameters.
We employ simulation results for $m_{\pi}$ 
obtained from the pion correlation
function at $L=28$.
The fit results are shown by dashed lines 
in Fig.~\ref{fig:ene_308} and Fig.~\ref{fig:ene_30415}; 
they agree quite well with the simulation results.
We note that the difference in volume dependence is caused by 
the sign of $C_{\pi}(p)$ between the unstable and stable cases.

\subsection{Volume dependence of $E_{\pi\pi}$}

To understand the volume dependence of $\pi\pi$ energy,
we use the scattering length $a_0$
that relates to the difference of $E_{\pi\pi}$ from $2m_{\pi}$
through L\"{u}scher's formula~\cite{Lu} given by 
$
E_{\pi\pi}-2m_{\pi}
= -\frac{4\pi a_0}{m_{\pi}L^3}
\left(1+c_1\frac{a_0}{L}+c_2\frac{a_0^2}{L^2}\right),
$
with $c_1 = -2.837297,\ c_2 = 6.375183$.
We employ the $\pi\pi$ energy obtained with diagonalization. 
The results are presented in Table~\ref{tab:scat-pert}.
The scattering length exhibits an opposite sign between the two cases.


We calculate the perturbative scattering length
from the perturbative phase shift~\cite{O4} and find
$
a_0 = \frac{g_R m_{\pi}}{96\pi m_{\sigma}^2}
\frac{7R^2+8}{R^2-4},
$
where $R=m_{\sigma}/m_{\pi}$.
The renormalized coupling $g_R$ is already determined in the 
$\sigma$ mass fit. Alternatively one may use the perturbative 
relation $g_R=3Z_{\pi}(m^2_{\sigma}-m^2_{\pi})/v^2$ where 
$m_{\pi}$, $Z_{\pi}$ and $v$ are taken from 
the simulation results for $L=28$.
We employ the $\sigma$ mass determined
from the $\sigma$ mass fit for $m_{\sigma}$.
We compile $a_0$ and $g_R$ from simulations and from perturbation 
theory in Table~\ref{tab:scat-pert}.

We find results from the simulation to be 
reasonably consistent with the perturbative results.
The sign of the perturbative scattering length 
depends only on whether the $\sigma$ particle
is unstable or stable.
From these signs the $\pi\pi$ energy in a finite volume is expected 
to increase or decrease in the unstable or stable cases,
respectively.
The behavior of the $\pi\pi$ energy also 
agrees with our simulation results.

\begin{table}[t!]
\begin{tabular}{lcccc}\hline
&\multicolumn{2}{c}{unstable}&\multicolumn{2}{c}{stable}\\ \hline
&\hspace{-0.3cm}$a_{0}$&\hspace{-0.3cm}$g_R$
&\hspace{-0.3cm}$a_{0}$&\hspace{-0.3cm}$g_R$\\ \hline
(S) diag.                &\hspace{-0.3cm}0.289(9)   &
                        &\hspace{-0.3cm}$-2.49(19)$&\\ 
(P) $g_R$(fit)           &\hspace{-0.3cm}0.383(3)   &\hspace{-0.3cm}14(1)  
                        &\hspace{-0.3cm}$-1.98(23)$&\hspace{-0.3cm}9(1)   \\ 
(P) $g_R({\mathrm def.})$&\hspace{-0.3cm}0.496(1)   &\hspace{-0.3cm}18.9(2)
                        &\hspace{-0.3cm}$-3.62(11)$&\hspace{-0.3cm}17.8(1)\\
\hline
\end{tabular}
\caption{
Scattering lengths
obtained from simulation (S) and perturbative formula (P).
\label{tab:scat-pert}
}
\vspace{-.6cm}
\end{table}

\section{Conclusion}
We have demonstrated that the maximum entropy method is a 
simple technique for decomposing the states
in the unstable particle correlation function
and determining their energies.
In future it will be interesting and challenging to apply this method to
studies of the $\rho$ meson decay and the $\sigma$ particle in QCD.

\vspace{2mm}
This work is supported in part by Grants-in-Aid of the Ministry of Education 
No.~12740133.

%

%
%


\begin{thebibliography}{l}




\bibitem{my}T.~Yamazaki {\it et al.} (CP-PACS Collaboration), 
Phys.\ Rev.\ 
{\bf D65} 014501 (2002).

\bibitem{O4}M.~G\"ockeler, H.A.~Kastrup, J.~Westphalen,
and F.~Zimmermann,
Nucl.\ Phys.\ 
{\bf B425} (1994) 413.

\bibitem{myO4}T.~Yamazaki and N.~Ishizuka, in preparation. 

\bibitem{ph1}M.~L\"uscher and U.~Wolff,
Nucl.\ Phys.\ 
{\bf B339} (1990) 222.

\bibitem{Lu}M.~L\"uscher,
Commun.\ Math.\ Phys.\ 
{\bf 105} (1986) 153.



\end{thebibliography}
\end{document}